\documentclass[showpacs,twocolumn,preprintnumbers,prl,amssymb]{revtex4}
\usepackage{graphicx} 
\usepackage{dcolumn}
\usepackage{bm}
\usepackage{amsmath}
\usepackage{amsfonts}
\usepackage{amssymb}
\usepackage{ulem}
\usepackage{color}

\begin{document}
\title{Polaron Transport in Organic Crystals:\\
Temperature Tuning of Disorder Effects}

\author{Frank Ortmann$^{1,2}$}
\author{Stephan Roche$^{2,3}$}
\affiliation{$^1$CEA, INAC, SPRAM, GT, 17 rue des Martyrs, 38054 Grenoble
Cedex 9, France}
\affiliation{$^2$ CIN2 (ICN-CSIC) and Universitat Aut\'{o}noma de Barcelona, Catalan Institute
 of Nanotechnology, Campus UAB, 08193 Bellaterra (Barcelona), Spain} 
\affiliation{$^3$ ICREA, Instituci\'{o} Catalana de Recerca i Estudis Avan\c{c}ats, 08070 
Barcelona, Spain}

\begin{abstract}
We explore polaronic quantum transport
in three-dimensional models of disordered organic crystals
with strong coupling between electronic and vibrational
degrees of freedom. By studying the polaron dynamics
in a static disorder environment, temperature dependent
mobilities are extracted and found to exhibit
different fingerprints depending on the strength of the
disorder potential. At low temperatures and for
strong enough disorder, coherence effects induce weak
localization of polarons. These effects are reduced with
increasing temperature (thermal disorder) resulting in
mobility increase. However at a transition temperature,
phonon-assisted contributions driven by polaron-phonon
scattering prevail, provoking a downturn of the
mobility. The results provide an alternative scenario to
discuss controversial experimental features in molecular
crystals.
\end{abstract}

\maketitle

The microscopic understanding of charge transport in organic materials remains a tantalizing 
challenge after more than two decades of theoretical and experimental efforts~\cite{RMP, Bredas}.
 An established fact is however that transport theories developed for conventional inorganic semiconductors fail to properly describe organic matter (or narrow-band systems in general) in its full complexity. 
One of the main reasons lies in the strong 
electron-phonon coupling which roots high technical and conceptual hurdles, demanding for the 
advancement of concepts and quantum transport methods able to cope with entangled 
effects of structural/chemical and thermal disorders. 
For strongly disordered (and soft) organic structures (such as discotic liquid crystals, DNA or 
microcrystalline polymer semiconductors), multiscale approaches combining classical molecular dynamics with 
quantum transport simulations offer an interesting perspective~\cite{MDT,MDT2}. On the other hand, 
since the seminal work of Su, Schrieffer, and Heeger\cite{SSH_RMP}, it is recognized that 
the electron transport in $\pi$-conjugated materials encompass various complicated transport mechanisms 
including band conduction or hopping processes which can be well captured by the concepts of 
polarons.

After early works by Holstein \cite{Holstein} and Conwell \cite{Conwell},
theoretical progress in recent years has been achieved in the 
exploration of polaronic phenomena, including the description of 
Fr\"{o}hlich polarons at the dielectric/organics interface in organic-field effect transistors  
(accounting for Coulomb interaction and screening phenomena) \cite{Fratini}, as well as the 
simulation of non-adiabatic polaron motion\cite{Stafstrom}, inelastic effects on coherent 
polaron-like motion~\cite{Ness}, or the revision of dynamic localization concepts in 
one-dimensional models of organic materials~\cite{troisiPRB2010}.  
Additionally, to investigate polaron motion in higher dimensional (3D) systems,
 a generalized transport methodology has been proposed using a mixed Holstein-Peierls model~\cite{Bobbert} in the Kubo framework.  This latter work has provided some qualitative understanding of high temperature-dependent charge mobility of ultrapure organic crystals (including anisotropic effects), but it does not explain the sign reversal of the bulk mobility versus temperature derivative ($\frac{\partial\mu}{\partial T}$)  in the lower temperature regime~\cite{EXP1} which still stands as a debated issue. Indeed, as observed by Karl and coworkers in $\alpha$-perylene single crystalline thin films, charge mobilities can increase from 1 cm$^2$(Vs)$^{-1}$ (10K) to around 100 cm$^2$(Vs)$^{-1}$ at 30 K,  followed by a power-law decay up to room temperature~\cite{EXP1}. A similar mobility behavior has been also more recently reported in rubrene crystals-based field effect transistors~\cite{EXP2} and 
in modified pentacene polycrystalline films with large domain sizes of over 100nm \cite{EXP3}.  
Notwithstanding repeated analysis, the underlying physical mechanism driving such temperature-dependent 
mobility remains elusive~\cite{KlocGirlando}.
Moreover, the reported clear evidences for band dispersion (as large as 400 meV) in high 
purity rubrene single crystals \cite{JAP} support the scenario of polaron band transport 
\cite{troisiPRB2010,OrtmannNJP,Zuppiroli}, although the question of quantum interference effects  \cite{LOC}
remains to be investigated and quantified in 3D disordered organic crystals.

In this paper, we use a real-space order N Kubo methodology to theoretically explore polaron transport 
in three-dimensional disordered organic crystals, taking into account elastic scattering, 
as well as decoherence effects.  
Decoherence is introduced at a phenomenological level, 
assuming that the coherent polaron motion suffers from additional stochastic dephasing events, 
driven by thermal vibrations which play an important role in the low-$T$ regime~\cite{Schmit}. 
Additionally, incoherent contributions 
of polaron motion (driven by polaron-phonon scattering~\cite{Bobbert}) are
found to dominate the high-$T$ regime. The obtained temperature-dependent 
mobility fingerprints are reminescent of experimental features in fairly ordered organic materials~\cite{EXP1,EXP2,EXP3}.

\begin{figure}[t]
\begin{center}
\leavevmode
\includegraphics[scale=0.9]{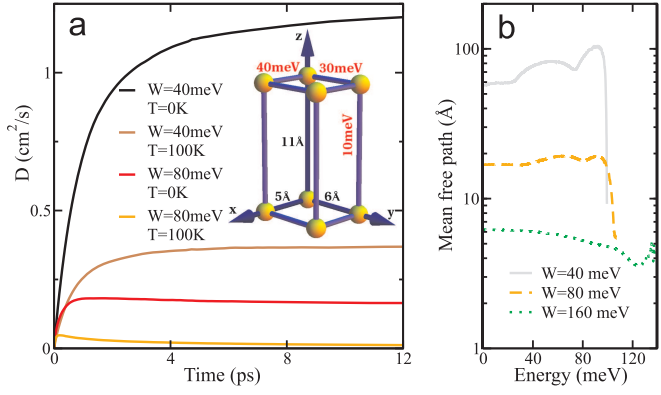}
\caption{(color online) (a) Typical time dependent diffusion coefficient for 
various $W$  and temperature $T$. Inset: 3D crystal structure 
with system parameters $\varepsilon_{mn}$ and $R_{mn}$ for
nearest neighbors. (b) Polaron mean free path versus energy for various  disorder strengths. Only positive energies are shown because of symmetry.}
\label{fig1}
\end{center}
\end{figure}

The present theoretical approach uses a polaronic tight-binding
(TB) Hamiltonian and orthorhombic symmetry (see inset of Fig. \ref{fig1}), as found in rubrene,
which is defined by $H=H_P+H_W$ where 
$H_P=\sum_{m\ne n}\widetilde{\varepsilon}_{mn}a^\dag_ma_n+\sum_\lambda \hbar\omega_{\lambda}b^\dag_\lambda b_\lambda$   
describes polarons and phonons. It can be obtained from
a polaronic transformation (Lang-Firsov)\cite{Bobbert} of a generic model for 
disordered polaronic materials.\cite{Fehske} In the present study, we use
the concept of a single effective vibrational mode with energy $\omega_{\lambda}$=12 meV, 
and we use as a typical dimensionless electron-phonon coupling $g_{\lambda}=0.7$
inspired by available {\it ab initio} data. The effective mode used here is
representative of the many low-frequency intermolecular
modes which are present in organic materials.
The temperature dependent renormalized transfer integrals $\widetilde{\varepsilon}_{mn}$
are used following the full dressing approximation \cite{Bobbert} ($N_\lambda$ is the Bose-Einstein function),
and are given by $\widetilde{\varepsilon}_{mn}=\varepsilon_{mn}\exp[-\sum_{\lambda}(1+2N_{\lambda})g^2_{\lambda}]$
with $\varepsilon_{mn}$ the electronic transfer integrals (given in Fig. \ref{fig1}-inset).
$H_P$ gives rise to a gapless cosine-type temperature dependent
and anisotropic band structure in the 3D Brillouin
zone, regarded as a typical case for organic crystals \cite{PentCoupling}.
Finally, $H_W=\sum_{m}\varepsilon_m a^\dag_ma_m$ 
tunes the strength of an additional elastic (Anderson-type) disorder potential~\cite{Zuppiroli}
with onsite energies taken at random within
$\varepsilon_m\in[-W/2,W/2]$ with an average modulus of $\varepsilon_m$ which is $W/4$. 
Such uncorrelated disorder mimics
impurities and defects as a first approximation. \cite{Zuppiroli}
A refined analysis could be achieved by using 
{\it first principles} calculations and the precise crystal symmetry
(see e.g. \cite{FirstPrinciplesParameters2} for details). 

The TB model is implemented into a Kubo transport framework
which is based on the theory developed in Ref. \cite{PolaronKubo}. 
This theory is generalized in the present work to investigate 
effects of static disorder. Due to the splitting of the mobility into
two contributions (coherent and incoherent) obtained in Ref. \cite{PolaronKubo}, we proceed
here by employing two complementary coherent and incoherent Kubo transport 
methodologies to explore the whole temperature range. The first one allows a real-space description 
of polaron propagation in the coherent regime in the superimposed disordered (static) potential 
using the Kubo-Greenwood approach~\cite{MKRT}. 
By following the polaron dynamics, the elastic polaron mean free path 
as well as the weak-localization corrections to the conductivity in the low-temperature regime 
can be computed. 
Second, the incoherent (phonon-assisted) contribution of polaron transport is 
conveyed by polaron-phonon scattering. It is calculated based on the 
theory developed in Ref~\cite{PolaronKubo}, 
which is here generalized to include static disorder effects (details below). 
Both contributions (coherent+incoherent) are computed separately and add to evaluate the total polaron mobility.
We note that there is no further parameter to tune the relative amplitudes of both.
They are fixed by material parameters and the impact of disorder.

From the study of the coherent polaron wave-packet dynamics, the elastic mean free path 
$\ell_{e}(E)$ can be first derived from the time dependence of the diffusion 
coefficient $D(E,t)$ which writes
$$\frac{1}{t} \frac{ \sum_l \langle \Psi_l \lvert \delta(E-H) \big{(} U^{\dag}(t) x U(t) - x \big{)}^2 \lvert \Psi_l\rangle} { \sum_l \langle \Psi_l \lvert \delta(E-H) \lvert \Psi_l \rangle },$$
taking $x$ as the position operator along the transport direction,
$\delta(E-H)$ the spectral measure operator the trace of which gives the
total density of states,
and $\lvert \Psi_l\rangle$ a set of random-phase states (for details see Ref. ~\cite{MKRT}). 
In this approach $D(t)$ is derived from to the time evolution of $\langle(x(t)-x(0))^2\rangle$ 
driven by the operator $\hat{U}(t)=\Pi_{n=1}^{N_t}\exp(iH\Delta t/\hbar)$ with 
$\Delta t$ the chosen time step. The calculations are 
performed for total elapsed computational times of $t_{\text{max}}=61$ ps. 
The system size is at least $(0.24\times0.22\times0.14)\mu \text{m}^{3}$ and periodic 
boundary conditions are applied. Based on this, the coherent part of the Kubo 
conductivity is computed using $\sigma(E, t=N_t\Delta t)=e^{2}_0\rho(E)D(E,t)/2$, 
with $\rho(E)$ the total density of states. We further derive the carrier mobility 
$$\mu^{\text{(coh)}}=\frac{\Omega}{e_0ck_BT}\int dE \sigma(E,t)f_{FD}(E)[1-f_{FD}(E)]$$ 
with $f_{FD}(E)$ being the Fermi-Dirac function, $c$ the total charge density fixed 
to $10^{-3}$, and $\Omega$ the volume of the unit cell.
The chosen time $t$ is fixed by the temperature dependent decoherence time $\tau_{\varphi}$.

By introducing static disorder, polaron wavepackets are elastically scattered and suffer from multiple scattering events and quantum interferences. Elastic scattering first results in a diffusive regime, characterized by the polaron elastic mean free path $\ell_{e}=D^{\text{max}}/2\widetilde{v}$ ($D^{\text{max}}$ is the maximum diffusion coefficient and $\widetilde{v}$ the polaron velocity). Fig.~\ref{fig1} (b) shows $\ell_{e}$ along the transport direction, which follows an approximate $\ell_{e}\propto W^{-2}$ behavior for small $W$ (in agreement with the Fermi golden rule).  The obtained values for $\ell_{e}$ are well below the total system size.
When $\ell_{e}$ becomes close to the lattice constant, 
it however ceases to be a suitable characteristic length as the system proceeds to 
the strong localization regime. In that limit the localization length $\xi$ might 
be used to further characterize localized states. 
As seen in Fig.1(b), the case of $W=160$ meV may be regarded (in our model) 
as a transition point. One also notes that the polaron bandwidth varies with temperature (renormalization of transfer integrals). As a result, for a given static disorder potential, backscattering efficiency will increase with temperature. 

Beyond the diffusive regime, quantum interference effects 
(QIE) produce an increase of the resistance (weak localization).
Strong (coherent) localization effects due to disorder in organic crystals have been studied in one 
and two-dimensional models~\cite{Zuppiroli}, for which all states are localized irrespective to the 
disorder strength. In three-dimensional models, the validity of band type conduction is known 
to be more robust with the existence of a metal-insulator transition, that can be tuned by the strength 
of elastic disorder~\cite{LOC}. 

\begin{figure}[t]
\begin{center}
\leavevmode
\includegraphics[angle=-90]{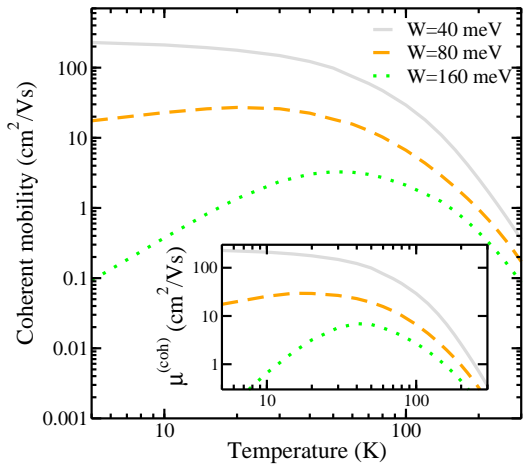}
\caption{(color online) Coherent part of the carrier mobility for several disorder strengths $W$ and using a decoherence model with $\alpha=2$ (see text). 
Inset: same information using $\alpha=3$.}
\label{fig2}
\end{center}
\end{figure}

In general, quantum coherence (and localization effects) will be reduced by decoherence mechanisms~\cite{LOC}.  Decoherence mechanisms for polaron transport could be driven by thermal disorders including phonon-phonon, electron-electron as well as electron-phonon scattering processes. 
This thermal disorder includes stochastic scattering
events which relate the selected coherent polaron
with other phonon degrees of freedom. We introduce
phenomenologically a coherence time which tunes the
weak localization correction, and decreases with temperature
following a power law. Although there is no
available microscopic theory of decoherence in inorganic
(or organic) materials, a power-law dependence of the
decoherence time $\tau_{\varphi}\propto 1/T^{\alpha}$ has been derived, 
with $\alpha$ only depending on the transport dimensionality for 
electron-electron interaction induced dephasing~\cite{LOC} 
(for temperatures below 1K), in contrast to the case 
of electron-phonon driven decoherence effects for which $\alpha\simeq 2-4$ 
(material dependent)~\cite{Schmit}. We here assume such a power law for $\tau_{\varphi}$
to be generic, but we take into account some variability in the $T$-dependent 
tuning of QIE by considering different values for 
${\alpha}$. We also assume that, at 5K, $\tau_{\varphi}$ corresponds to 
the maximum computed time ($\tau_\varphi=t_{\text{max}}$) and introduce a 
lower limit for $\tau_\varphi$ which ensures a seamless transition to 
the semiclassical limit at elevated $T$.  

Fig.~\ref{fig2} shows $\mu^{\text {(coh)}}(T)$ for various disorder strengths $W$. 
For values as low as 40 meV (corresponding to elastic mean free paths in the 
order of 6-10 nm, Fig.~\ref{fig1} (b)),
no weak localization is observed within the reach of our maximum computed time.
This can be appreciated by the absence of a time-dependent decay of the diffusion coefficient
(see Fig.~\ref{fig1} (a)). As a result, the low-temperature $\mu^{\text {(coh)}}(T)$
is seen to saturate to its semiclassical value on the order of 200 cm$^2$(Vs)$^{-1}$
for the chosen parameters. 
In contrast, for a static disorder strength such as $W$=80 meV and $W$=160 meV
(corresponding to $\ell_{e}\sim 2$nm  and $\ell_{e}\sim 6$\AA, respectively) substantial contributions of QIE are obtained, as witnessed by the
time-dependent decay of the diffusion coefficient (Fig.1(a)), and result in a marked decrease of $\mu^{\text {(coh)}}$ at the lowest temperatures. 
By increasing $W$ by a factor of 2 (or 4), $\mu^{\text {(coh)}}(T)$ 
is reduced by 1 (or 3) orders of magnitude. The temperature-dependent 
decoherence effects then yield a positive $\frac{\partial\mu}{\partial T}$ (Fig.~\ref{fig2})
which is independent of the exact details of the decoherence strength introduced phenomenologically.
Such a behaviour can only be observed if the temperature-induced reduction of QIE dominates the
band narrowing effect which is not observed for the negative $\frac{\partial\mu}{\partial T}$
found at higher temperatures in Fig. ~\ref{fig2}.
Importantly, in contrast to common belief that a mobility increase with temperature implies 
a phonon assisted transport regime, 
we obtain a positive $\frac{\partial\mu}{\partial T}$ for coherent motion.
This behavior can be expected to be relatively insensitive to the strength of the 
electron-phonon coupling or the efficiency of the phonon dressing of carriers
assumed in the polaron Hamiltonian.

\begin{figure}[t]
\begin{center}
\leavevmode
\includegraphics[angle=-90]{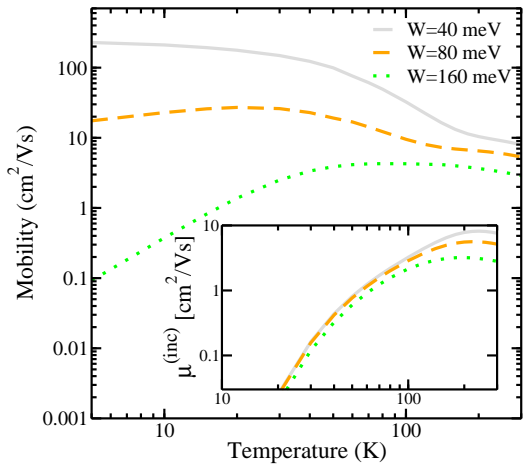}
\caption{(color online) Temperature-dependence of the total mobility (main frame) and incoherent part (inset) for several disorder strengths $W$ and decoherence effects 
with $\alpha=2$.}
\label{fig3}
\end{center}
\end{figure}

At elevated temperatures, the phonon-assisted incoherent contribution 
of polaron transport (driven by phonon-polaron scattering) may prevail 
over the coherent one. By generalizing Eq. (43) of Ref. ~\cite{PolaronKubo} 
(using the density of states $\rho(E)$ of the disordered system), 
we compute the incoherent part of the polaron mobility
$\mu^{\text{(inc)}}=\frac{e_0\Omega^2}{2c\hbar^2k_BT}\sum_mR_{mn}^2\widetilde{\varepsilon}_{mn}^2\int dE_1dE_2\rho(E_1)\rho(E_2)f_{FD}(E_1)[1-f_{FD}(E_2)]\int_{-\infty}^{\infty}dt e^{it(E_1-E_2)/\hbar}\{\exp[2\Phi_\lambda(t)g_\lambda^2]-1\}$ 
in transport direction
with $\Phi_\lambda(t)=N_\lambda e^{i\omega_\lambda t}+(1+N_\lambda)e^{-i\omega_\lambda t}$
and display its temperature dependence in Fig. \ref{fig3}.~\cite{note}  We observe an 
activation behaviour of the phonon-assisted transport which is 
followed by a saturation regime at about 200K and a weak decrease 
up to room temperature. Although reminescent of the experimental
data reported recently~\cite{EXP1,EXP2,EXP3}, we discuss below why
phonon-assisted transport is unlikely to explain the mobility downturn at low $T$.
We plot in Fig. \ref{fig3} (main frame) the total carrier mobility 
$\mu^{\text {(coh)}}(T)+\mu^{\text {(inc)}}(T)$, which is the
measurable quantity in experiments. For all temperatures one observes a 
disorder dependence of the mobility which is weaker for the high-$T$ regime. However, its 
origin in this regime is very different from the strong quantum interferences 
described in Fig.~\ref{fig2} at low-$T$.
Importantly, we find that even for strong disorder the weak localization 
regime of coherent motion dominates the low-temperature regime of 
the total carrier mobility, resulting in $\frac{\partial\mu}{\partial T}>0$
for strong enough disorder (not phonon-assisted). 
On the other hand the mobility decrease ($\frac{\partial\mu}{\partial T}<0$) 
at elevated $T$ is mostly driven by phonon-assisted polaron motion,
i.e. increasing contribution of carrier-phonon scattering, 
in contrast to the temperature-activation concept commonly 
associated with phonon-assisted transport.

{\it Conclusion}.-We have studied a quantum-based polaron transport scenario
and found that quantum interferences can play an important role for 
understanding temperature-dependent transport features. 
Such scenario is consistent with experimental trends reported in clean organic 
crystals~\cite{EXP1,EXP2,EXP3}, and the obtained results are not within the reach
of restricted semiclassical transport approaches. Although the absolute values of charge mobilities
are comparable to what is measured experimentally, a systematic comparison with
 experimental data, including transport anisotropy, would demand 
the use of first principles calculations for the material parameters.
This work is supported by a Marie Curie Intra European
Fellowship within the 7th European Community Framework Programme.


\begin{thebibliography}{50}
\bibitem{note} A phenomenological parameter $\tau$  which fixes
the absolute mobility values in Ref. 25 is no longer necessary and omitted here.
\end{thebibliography}
\end{document}